\documentclass[preprint,aps,floats,showpacs,nofootinbib]{revtex4}
\usepackage{graphicx}
\usepackage{epsfig}
\usepackage{pstricks}
\usepackage{pst-coil}
\usepackage{amsmath}
\def\be{\begin{equation}}
\def\ee{\end{equation}}
\def\bea{\begin{eqnarray}}
\def\eea{\end{eqnarray}}
\begin{document}
~
\begin{flushright}UCRHEP-T423\\November 2006
\end{flushright}
\vskip 1.0cm
\title{\Large \bf Low-Energy Thermal Leptogenesis\\ in 
an Extended NMSSM Model}
\author{Ernest Ma}
\affiliation{Physics and Astronomy Department, University of California, 
Riverside, CA 92521, USA}
\author{Narendra Sahu}
\email{narendra@prl.res.in}
\author{Utpal Sarkar}
\affiliation{Theory Division, Physical Research Laboratory,
Navrangpura, Ahmedabad, 380 009, India}

\begin{abstract}\
\noindent
Thermal leptogenesis in the canonical seesaw model in supersymmetry suffers 
from the incompatibility of a generic lower bound on the mass 
scale of the lightest right-handed neutrino and the upper bound 
on the reheating temperature of the Universe after inflation. 
This is resolved by adding an extra singlet superfield, with 
a discrete $Z_2$ symmetry, to the NMSSM (Next to Minimal Supersymmetric 
Standard Model).  This generic
mechanism is applicable to any supersymmetric model for lowering 
the scale of leptogenesis.
\end{abstract}
\maketitle
\newpage
\baselineskip 24pt
\section{Introduction}
\noindent
In the minimal standard model (SM), neutrinos are massless. However, 
small nonzero neutrino masses are required by the atmospheric
and solar neutrino experiments. A natural explanation for such 
tiny neutrino masses in the SM comes from an effective dimension-5 
operator~\cite{w79}
\begin{equation}
{\cal L}_\Lambda = \frac{f_{\alpha \beta}}{\Lambda} (\nu_\alpha \phi^0 - 
l_\alpha \phi^+)(\nu_\beta \phi^0 - l_\beta \phi^+) + H.c.,
\end{equation}
where $(\nu_\alpha,l_\alpha),~\alpha=e,\mu,\tau$ are the usual left-handed 
lepton doublets transforming as $(2,-1/2)$ under the standard electroweak 
$SU(2)_L \times U(1)_Y$ gauge group and $(\phi^+,\phi^0) \sim (2,1/2)$ is 
the usual Higgs doublet of the SM. There are three realizations of this 
operator~\cite{m98}, the most popular one being the canonical 
seesaw~\cite{seesaw} mechanism which adds three singlet heavy neutral 
fermions $N_i,~i=1,2,3$ to the SM Lagrangian. The neutrino mass matrix 
is then given by
\begin{equation}
{\cal M}_\nu = - {\cal M}_D {\cal M}_N^{-1} {\cal M}_D^T\,
\label{neu_mass}
\end{equation}
where ${\cal M}_D$ is the $3 \times 3$ Dirac mass matrix linking
$\nu_\alpha$ with $N_i$ through the Yukawa interactions $h_{\alpha i}
(\nu_\alpha \phi^0 -\ell_\alpha \phi^+) N_i$. 

The Majorana masses of $N_i$ violate lepton number by 
two units. Therefore, in the early Universe, a net lepton asymmetry may be 
generated~\cite{fy86} through the out-of-equilibrium decay of the 
lightest $N_i$ (call it $N_1$). The generated lepton asymmetry then 
gets converted into a baryon asymmetry through the interactions of 
the SM sphalerons~\cite{kr85} which conserve $B-L$, but violate $B+L$, 
where $B$ and $L$ are baryon and lepton number respectively. The
existence of $N_i$ explains thus at the same time why both neutrino
masses as well as the observed baryon asymmetry of the Universe (BAU)
are nonzero and small.

In supersymmetric theories the reheating temperature ($T_h$) 
following inflation is likely to be rather low~\cite{gravitino,kkm}. 
Although some models~\cite{am} may allow a higher reheating temperature, 
in the conventional models $T_h$ is bounded strongly from above by 
the possible overproduction of gravitinos. 
On the other hand, in the simplest version of the seesaw mechanism the 
condition for thermal leptogenesis requires that the mass 
$M_1$ of the lightest right-handed neutrino $N_1$ should be
much higher than $T_h$, assuming that $N_1$ contributes to the 
left-handed neutrino masses dominantly. Since inflation would erase 
any pre-existing lepton asymmetry, the asymmetry generated by $N_1$ 
after inflation would be highly suppressed by its small number density 
and hence this mechanism will fail to explain the BAU.

To avoid this problem, several ideas have been discussed in the
literature~\cite{ideas, riotoetal,tevscale_lep}. An attractive 
scenario is the extended seesaw mechanism~\cite{ma_sahu_sarkar.06, 
valle&sarkar.06}. In this paper we follow the same scheme and work 
with the canonical 
seesaw mechanism (SM plus three $N_i$) in the Next to Minimal Supersymmetric
Standard Model (NMSSM), which has an extra singlet superfield $\chi$. 
To distinguish $N_i$ from $\chi$, an exactly conserved $Z_2$ discrete 
symmetry is imposed, corresponding to $(-1)^L$. We then add an extra 
heavy singlet superfield $S$, together with a softly broken discrete 
symmetry $Z_2^\prime$, under which $S$ is odd and all others are even. 
As a result, the production and decay channels
of $S$ are different, and the out-of-equilibrium decay of $S$
can take place much below the mass scale of the lightest right-handed 
neutrino $N_1$. The lower bound on the mass scale of $S$ can then 
be compatible with the upper bound on the reheating temperature after 
inflation.

The rest of this paper is arranged as follows.  In section II
we review the canonical leptogenesis and briefly recall the
Davidson-Ibarra (DI) bound on the mass scale of $N_1$. In
section III we discuss an extended seesaw model by introducing an
additional heavy singlet fermion $S$ of mass less than that of
$N_1$. In section IV we discuss how the thermal-leptogenesis
constraint on the mass scale of $S$ can be lowered in comparison
with the mass scale of $N_1$. In section V we solve the
required Boltzmann equations numerically and show how the
low mass scale of $S$ is compatible with thermal leptogenesis.
In section VI we state our conclusions.

\section{Canonical leptogenesis and DI bound}
\noindent
In canonical leptogenesis the lightest right-handed neutrino
$N_1$ decays into either $\ell^- \phi^+$ and $\nu \phi^0$, or $\ell^+
\phi^-$ and $\bar \nu \bar \phi^0$. Thus a CP asymmetry can be
established from the interference of the tree-level amplitudes
with the one-loop vertex~\cite{fy86} and self-energy corrections~\cite{fps}.
A net lepton asymmetry arises when the decay rate
\begin{equation}
\Gamma_1= \frac{1}{8\pi v^2}\tilde{m}_1 M_1^2
\label{gamma_n1}
\end{equation}
fails to compete with the expansion rate of the Universe
\begin{equation}
H(T) = 1.66 g_*^{1/2} \frac{T^2}{M_{pl}}
\label{hubble}
\end{equation}
at $T \sim M_1$, where $\tilde{m}_1=(m_D^\dagger m_D)_{11}/M_1$ is
the effective neutrino mass parameter, $g_* \simeq 228 $ is the
effective number of relativistic degrees of freedom in the MSSM
and $M_{pl} = 1.2 \times 10^{19}$ GeV. This means that a
\underline{upper} bound on $\tilde{m}_1$ may be established by first
considering the out-of-equilibrium condition $H(T=M_1) > \Gamma_1$
which gives
\begin{equation}
\tilde{m}_1<1.6\times 10^{-3} eV\,.
\label{upper_bound}
\end{equation}
However, for $\tilde{m}_1>10^{-3} eV$ a reduced lepton asymmetry may
still be generated, depending on the details of the Boltzmann
equations which quantify the deviation from equilibrium of the
process in question.

Assuming a normal mass hierarchy in the right handed neutrino mass
spectrum the CP asymmetry $\epsilon_1$ is given by
\begin{equation}
\epsilon_1 \simeq -\frac{3}{8\pi v^2} \left( \frac{M_1}{M_2} \right)
\frac{Im [(m_D^\dagger m_D)_{12}]^2}{(m_D^\dagger m_D)_{11}}\,.
\label{epsilon_1}
\end{equation}
                                                                                
The baryon-to-photon ratio of number densities has been measured
\cite{eta_b} with precision, i.e.
\begin{equation}
\eta_B \equiv \frac{n_B}{n_\gamma} = 6.1^{+0.3}_{-0.2}\times 10^{-10}.
\label{b-asym}
\end{equation}
To get the correct value of $\eta_B$, one needs $\epsilon_1$ to be
of order $10^{-6}$ to $10^{-7}$. However, using the DI upper bound on
the CP asymmetry parameter~\cite{di02}
\be
|\epsilon_1|\leq \frac{3M_1}{8\pi v^2}\sqrt{\Delta m_{atm}^2}
\label{cp-bound}
\ee
one can get a lower bound on the mass scale of the lightest right-handed
neutrino as
\begin{equation}
M_1\geq 2.9\times 10^9 GeV \left(\frac{\eta_B}{6.1\times 10^{-10}}
\right) \left(\frac{4\times 10^{-3}}{(n_{N_1}/s)
\delta}\right)\left(\frac{v}{174GeV}\right)^2\left(\frac{0.05eV}
{\sqrt{\Delta m_{atm}^2}}\right)\,,
\label{di-bound}
\end{equation}
where $v$ is the vacuum expectation value (vev) of the SM Higgs and it
is also assumed that the physical left handed neutrinos follow the
normal mass hierarchy. Since $T_h$ is not likely to exceed $10^9$ GeV,
this poses a problem for canonical leptogenesis. In order to overcome
this problem we consider an extended seesaw model as follows.

\section{The model for thermal leptogenesis below the DI bound}
\noindent
In a recent paper~\cite{ma_sahu_sarkar.06} we proposed a 
singlet mechanism to overcome the DI bound shown in 
Eq. (\ref{di-bound}). We now present a realistic model where
this mechanism can be implemented. One more ingredient has been added 
to produce these singlet fields abundantly in a thermal bath. 

We start with the NMSSM model, which includes a singlet 
superfield $\chi$ in addition to the usual particles of the
Minimal Supersymmetric Standard Model (MSSM). To implement the seesaw
mechanism, we also include three right-handed neutrinos $N_i,~i=1,2,3$. 
We then demonstrate how a minimal extension of this model may
admit a very low scale of leptogenesis, thus overcoming 
the gravitino problem.  We include another singlet superfield $S$ and 
impose a $Z_2 \times Z_2^\prime$ discrete symmetry.  Under $Z_2$, 
the lepton superfields $L_i, l^c_i, N_i, S$ are odd, whereas the Higgs 
superfields $\phi_{1,2}, \chi$ are even.  This corresponds to having 
an exactly conserved lepton number $(-1)^L$, or the usual $R-$parity of 
the MSSM.  Under $Z_2^\prime$, $S$ is odd and all others are even, 
but $Z_2^\prime$ is allowed to be broken softly.  
The most general superpotential invariant under $Z_2 \times Z_2^\prime$ 
is then given by
\begin{eqnarray}
W &=& h^e_{ij} L_i l_j^c \phi_1 + h_{ij} L_i N_j \phi_2 + \mu \phi_1 \phi_2
+ M_{ij} N_i N_j + M_\chi \chi \chi \nonumber \\ &&
+ \alpha \chi \chi \chi
+ \beta \chi \phi_1 \phi_2 + f_N \chi N_i N_j + M_S S S
+ f_S \chi S S  .
\end{eqnarray}
We do not discuss quarks nor other interactions of 
the NMSSM, which have been studied elsewhere. We deal 
only with neutrino masses and leptogenesis. 

We now break $Z_2^\prime$ softly and the only possible such term is 
$$ W_s = d_i N_i S ,$$
i.e. exactly as required to implement the singlet mechanism of 
ref.~\cite{ma_sahu_sarkar.06}. This allows 
$S$ to mix with $N_i$ to form a $7 \times 7$ mass matrix 
in the basis $[L_i ~~ N_i ~~ S]$, i.e.
\begin{equation}
\mathcal{M} = \begin{pmatrix}\bf{0} & m_D & 0\cr
m_D & M_N & d\cr
0 & d & M_S\end{pmatrix}
\label{mass-matrix}
\end{equation}
where $M_S=f_S\langle \chi \rangle$, and without loss of generality we 
choose $M_N$ to be diagonal with masses $M_{1,2,3}$.  For small 
$d_i/(M_i-M_S)$ as well as the usual seesaw assumption that the entries 
of $m_D$ are very small relative to $M_i$, the heavy masses are roughly 
given by
\begin{eqnarray}
M_{S'} & \simeq & M_S - \sum_i \frac{d_i^2}{M_i-M_S}, \nonumber\\ 
M_{N'_i} & \simeq & M_i + \frac{d^2}{M_i-M_S}, 
\end{eqnarray}
corresponding to the mass eigenstates $S'$ and $N'_i$ 
\begin{eqnarray}
S' &\simeq & S - \sum_i \frac{d_i}{M_i-M_S} N_i \nonumber\\
N'_i &\simeq & N_i + \frac{d_i}{M_i-M_S} S.
\end{eqnarray}
The light neutrino mass matrix is then
\be
(m_\nu)_{ij} \simeq -\sum_k (m_D)_{ik} \left( M_k + \frac{d_k^2}{M_k-M_S} 
\right)^{-1} (m_D)_{kj}.
\label{neutrino_mass}
\ee
In the limit $d_i \rightarrow 0$ we recover the neutrino masses as in 
the canonical seesaw mechanism.  We assume thus $M_{S'}\simeq M_S$ and 
$M_{N'_i}\simeq M_N$ in the following.

\section{Lepton asymmetry and lower bound on $M_S$}
In this model, the addition of $S$ allows the choice 
$M_S < M_1$. The induced couplings of $S$ to leptons are suppressed
by factors of $d_i/M_i$ compared to those of $N_i$. The decay rate
of $S$ is thus given by
\be
\Gamma_D^S =\frac{1}{8\pi v^2}M_S \sum_i
\left[ \tilde{m}_i M_i(d_i/M_i)^2\right]\,,
\label{gamma_D}
\ee
where we have assumed $M_s\ll M_i$ and the effective neutrino mass
parameter is defined as
\be
\tilde{m}_i=\frac{(m_D^\dagger m_D)_{ii}}{M_i}\,.
\label{effective-mass}
\ee
Assuming that $\frac{d_1}{M_1}=\frac{d_2}{M_2}=\frac{d_3}{M_3}$ and
$\tilde{m}_3 > \tilde{m}_2 > \tilde{m}_1$ the above Eq. (\ref{gamma_D})
can be rewritten as
\be
\Gamma_D^S \simeq \frac{1}{8\pi v^2}M_S \tilde{m}_3 M_3 (d_3/M_3)^2\,.
\label{final_gamma}
\ee
The out-of-equilibrium condition $\Gamma_D^S < H(T \sim M_S)$ is thus
suppressed by a factor
\begin{equation}
\eta=\left( \frac{d_3^2}{M_3 M_S} \right)\left( \frac{\tilde{m}_3}
{\tilde{m}_1}\right) \equiv \kappa \left( \frac{\tilde{m}_3}
{\tilde{m}_1}\right)\,
\label{suppression-factor}
\end{equation}
in comparison to $\Gamma_1/H(T\sim M_1)$ and hence can be satisfied at
a lower mass depending on the value of $d_3$. The value of $\tilde{m}_3$,
$\tilde{m}_1$ and $M_3$ can be approximately fixed from the low energy
neutrino oscillation data. So, the remaining free parameters are
$d_3$ and $M_S$ on which the suppression factor $\eta$ depends.
                                                                                
\begin{figure}[!thb]
\vskip 1.25in
\epsfxsize=120mm
\hskip -.95in
\centerline{\epsfbox{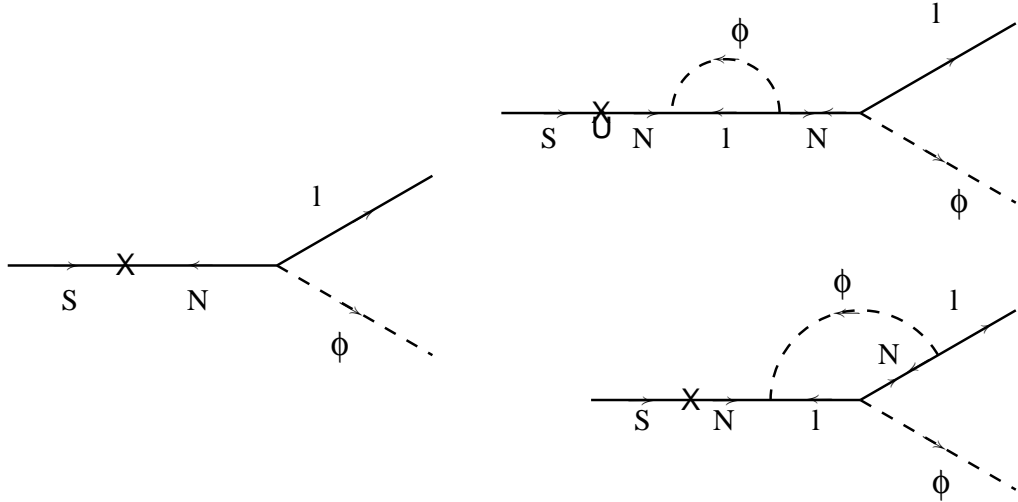}}
\vskip -.5in
\caption{Tree-level and one-loop (self-energy and
vertex) diagrams for $S$
decay, which interfere to generate a lepton asymmetry.  }
\label{lepfg}
\end{figure}
                                                                                
The CP asymmetry generated by the decays of $S$ comes from the interference
of the tree-level and one-loop diagrams of fig. (\ref{lepfg}). Both the
numerator and denominator of Eq. (\ref{epsilon_1}) are then suppressed
by the same $(d_i/M_i)^2$ factor, and we obtain
\begin{equation}
\epsilon_{S} \simeq -\frac{3}{8\pi v^2} \left( \frac{M_S}{M_2} \right)
\frac{Im [(m_D^\dagger m_D)_{12}]^2}{(m_D^\dagger m_D)_{11}}\,,
\label{epsilon_s}
\end{equation}
where we have assumed that $M_1\ll M_2\ll M_3$. Thus we see that the
CP asymmetry parameter is independent of the suppression parameter
$d_3$. Therefore, depending on the value of $\kappa$ the $L$-asymmetry
will saturate at different temperatures as implied by Eq.
(\ref{suppression-factor}). This is shown in Section V by numerically
solving the required Boltzmann equations. As demonstrated in
fig. (\ref{allowed_Ms}) the value of $M_S$ can be lowered much less
than the DI bound on $M_1$ by an appropriate choice of $d_3$. This
is because the low values of $M_S$ are not restricted by the low
energy neutrino oscillation data for $d_i\ll M_i$ as we have seen
from Eq. (12). Moreover the washout effects are suppressed for
low values of $(d_3/M_3)$. So, a successful lepton asymmetric
universe before the electroweak phase transition can be created
even for a TeV scale of $S$.

\section{Numerical estimation of lepton asymmetry}
\subsection{Production and decay of $S$}
\noindent
In this model $S$ is produced through the decay of $\chi$. The
corresponding Yukawa coupling $f_S$ can be as large as of order
unity. Therefore, $S$ can be brought to thermal equilibrium through
the scattering processes: $S \bar{S} \rightarrow S \bar{S}$,
$S\bar{S}\rightarrow \chi\chi^\dagger$ and $\chi S \rightarrow \chi S$.
Note that these processes never change the number density of $S$, but
they keep $S$ in kinetic equilibrium. The decay rate of $\chi$ can be
given by
\begin{equation}
\Gamma_D^\chi=\frac{f^\dagger f}{8\pi}M_\chi \left( 1-\frac{4M_S^2}
{M_\chi^2} \right)^{3/2} \frac{K_1\left( (M_\chi/M_S)z \right)}{K_2\left(
(M_\chi/M_S)z \right)}\,.
\label{decay_chi}
\end{equation}
where $(K_1/K_2)$ is the boost factor and $z=M_S/T$. Thus the inverse
decay of $\chi$ is given by $\Gamma_{ID}^{\chi}=(n_\chi^{eq}/n_S^{eq})
\Gamma_D^{\chi}$.
                                                                                
\begin{figure}[!thb]
\begin{center}
\epsfig{file=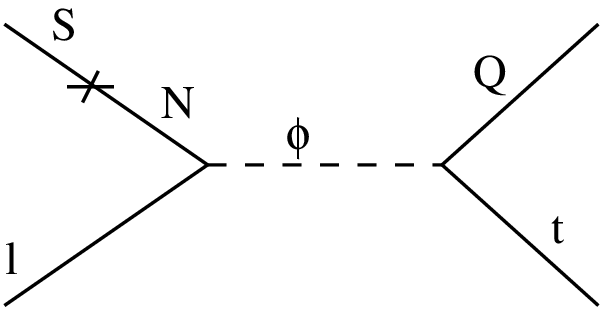, width=0.3\textwidth}
\epsfig{file=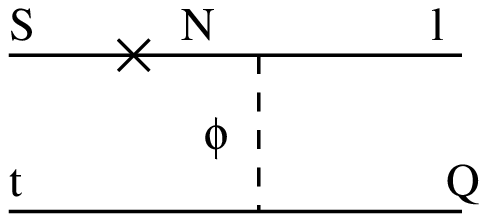, width=0.33\textwidth}
\end{center}
\caption{$\Delta L=\pm 1$ processes which deplete the number density of
$S$. These  processes also deplete the net lepton number density produced
through the decay channel.}
\label{delta_L=1}
\end{figure}

Once the $S$ particles are produced, they decay through the channel:
$S\rightarrow \ell \phi^\dagger, \bar{\ell} \phi$ as shown in 
fig. (\ref{lepfg}) which violates lepton number by two units. 
Apart from that, the other process
which depletes the number density of $S$ is $S\ell \rightarrow
\phi \rightarrow Q\bar{t}$. This is shown in fig. (\ref{delta_L=1}).
                                                                                
The subsequent decay of $S$, below its mass scale, then produces the
required baryon asymmetry through the leptogenesis route. However, an
exact lepton asymmetry can be estimated by solving the required
Boltzmann equations~\cite{boltzmann_papers}. It is useful to define
the Boltzmann equations in terms of the dimensionless variables
$Y_S=n_S/s$ and $Y_L=n_L/s$, where $Y_S$ is called the comoving
density of $S$ while $Y_L$ is the density of net lepton in a
comoving volume and
\be
s=\frac{2\pi^2}{45}g_* T^3
\label{entropy}
\ee
is the entropy density. The required Boltzmann equations are
given as
\be
\frac{dY_{S}}{dz}=-\left(Y_{S}-Y_S^{eq}\right) \left[ \frac{\Gamma_D^S}
{zH(z)}+\frac{\Gamma_s^S}{zH(z)} \right]
\label{S-evolution}
\ee
and
\be
\frac{dY_{L}}{dz}=\epsilon_S \frac{\Gamma_D^S}{zH(z)} \left( Y_{S}-Y_S^{eq}
\right)-\frac{\Gamma_W^\ell}{zH(z)}Y_L\,,
\label{L-evolution}
\ee
where $\Gamma_D^S$, $\Gamma_s^S$ and $\Gamma_W^\ell$ simultaneously
represent the decay, scattering and wash out rates that takepart in
establishing a net lepton asymmetry in a thermal plasma. The Hubble
parameter $H(z)$ is given by
\be
H(z)=\frac{H(M_S)}{z^2}~~~~~~{\rm with}~~~~H(M_S)=1.67 g_*^{1/2}
\frac{M_S^2}{M_{pl}}\,.
\label{hubble_mod}
\ee
In a relativistic frame the decay rate (\ref{final_gamma}) can be 
rewritten as
\be
\Gamma_D^S = \frac{1}{8\pi v^2}M_S \left( \frac{K_1(z)}{K_2(z)}\right)
\tilde{m}_3 M_3 (d_3/M_3)^2\,.
\label{final_decay}
\ee
                                                                                
The $\Gamma_s^S$ in Eq. (\ref{S-evolution}) represents the processes which
violate lepton number by one unit and is given by~\footnote{We have not
included the SUSY processes. It is shown that upon inclusion of those
processes the result doesn't change significantly~\cite{plumacher_npb}.}
\be
\Gamma_S=4\Gamma^{S}_{\phi,s}+2\Gamma^{S}_{\phi,t}\,.
\label{gamma_s}
\ee
\begin{figure}
\begin{center}
\epsfig{file=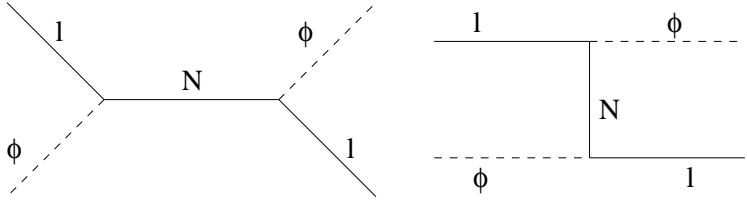, width=0.6\textwidth}
\end{center}
\caption{$\Delta L =\pm 2$ processes which deplete the number density
of net leptons.}
\label{washoutfig}
\end{figure}

The $\Gamma_W$ in Eq. (\ref{L-evolution}) represents the lepton number
violating processes by two units and is given by
\be
\Gamma_W=\frac{1}{2}\Gamma^{S}_{ID}+2\Gamma^l_{\phi,t}+2\Gamma^l_{\phi,s}
\left( \frac{Y_{N1}}{Y_{N1}^{eq}} \right)+2\Gamma^l_{N1}+2\Gamma^l_{N1,t}\,,
\label{washout}
\ee
where $\Gamma^S_{ID}=(n_S^{eq}/n_l^{eq})\Gamma^S_D$. In
Eqs. (\ref{gamma_s}) and (\ref{washout}) the $\Gamma$'s are defined
as $\Gamma^x_i=(\gamma_i/n_x^{eq})$ where $\gamma_i$ is the scattering
density. Note that the other $\Delta L =\pm 2$ processes: $ll\rightarrow
N_1 S N_1 \rightarrow \bar{\phi}\bar{\phi}$, and of course higher order
processes, which contribute to $\Gamma_W^\ell$ are suppressed in
comparison to the processes shown in fig. (\ref{washoutfig}).

\subsection{Solution of Boltzmann equations}
\begin{figure}
\begin{center}
\epsfig{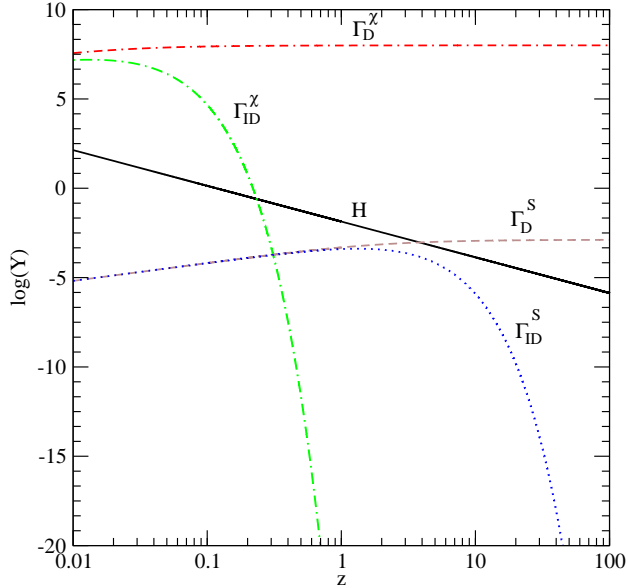}
\end{center}
\caption{The production and decay, and their inverse processes,
are compared with the Hubble expansion parameter $H$ for a typical
set of parameters. We have used $M_S=10^8$ GeV, $M_\chi=10^{10}$ GeV,
$M_3=10^{10} GeV$, $\tilde{m}_3=10^{-2}$ eV, $d_3=10^8$ GeV and
$f_S=0.5$.}
\label{decay_fig}
\end{figure}

\noindent
In fig. (\ref{decay_fig}) we have plotted the decay and inverse decay
of $\chi$ and $S$ against $z$. It is shown that the inverse decay of
$\ell+\phi^\dagger \rightarrow S$ is not sufficient to bring $S$ into
thermal equilibrium even if the suppression factor $d_3$ is as large
as $10^8$ GeV. On the contrary, the decay rate of $\chi$ is sufficiently
larger than the Hubble expansion parameter. Hence $S$ can be brought to
thermal equilibrium through the scattering process involving $\chi$ as 
long as $(M_\chi/M_S)\simeq O(10^{1-2})$. Therefore, at a temperature 
far above the mass scale of $S$ it is in
thermal equilibrium and hence a net lepton asymmetry in the thermal
plasma is zero. Below the mass scale of $S$ the lepton number violating
processes go out of thermal equilibrium and thus produce a net lepton
asymmetry dynamically. This is obtained by solving the Boltzmann
equations (\ref{S-evolution}) and (\ref{L-evolution}). We take the
following initial conditions:
\be
Y_S=Y^{eq}_S~~~~ {\rm and} ~~~~Y_L=0 ~~~~{\rm at}~~~~ z\rightarrow 0\,.
\label{initial-condition}
\ee
\begin{figure}
\begin{center}
\epsfig{file=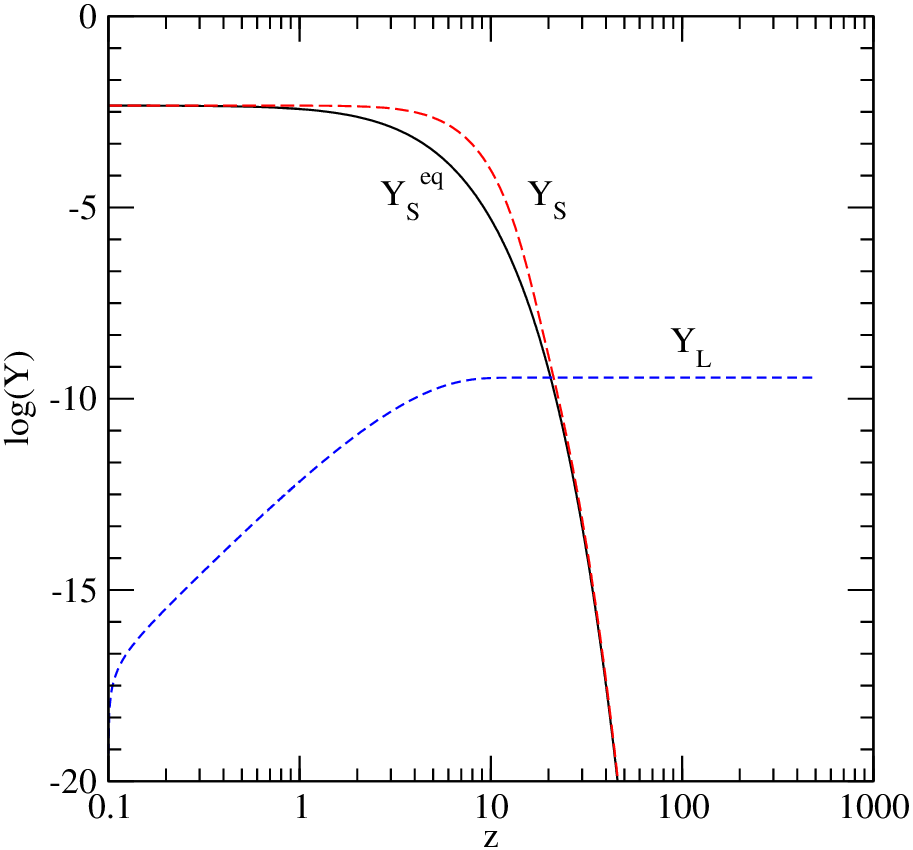, width=0.5\textwidth}
\end{center}
\caption{The evolution of $S$ is shown against $z$ with $M_S=10^8$ GeV,
$M_1=10^9$ GeV, $M_3=10^{10}$ GeV and $d_3=10^8$ GeV and the CP asymmetry
parameter is $\epsilon_S=10^{-7}$. }
\label{d38}
\end{figure}
\begin{figure}
\begin{center}
\epsfig{file=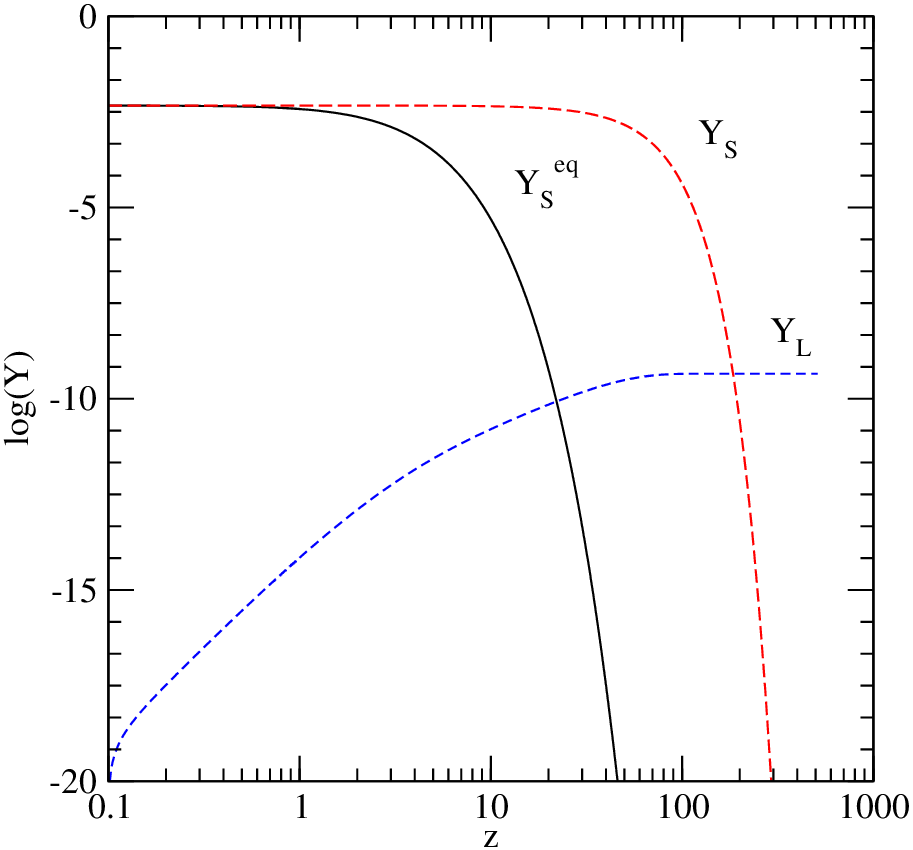, width=0.5\textwidth}
\end{center}
\caption{The evolution of $S$ is shown against $z$ with $M_S=10^8$ GeV,
$M_1=10^9$ GeV, $M_3=10^{10}$ GeV and $d_3=10^7$ GeV and the CP asymmetry
parameter is $\epsilon_S=10^{-7}$. }
\label{d37}
\end{figure}
\begin{figure}
\begin{center}
\epsfig{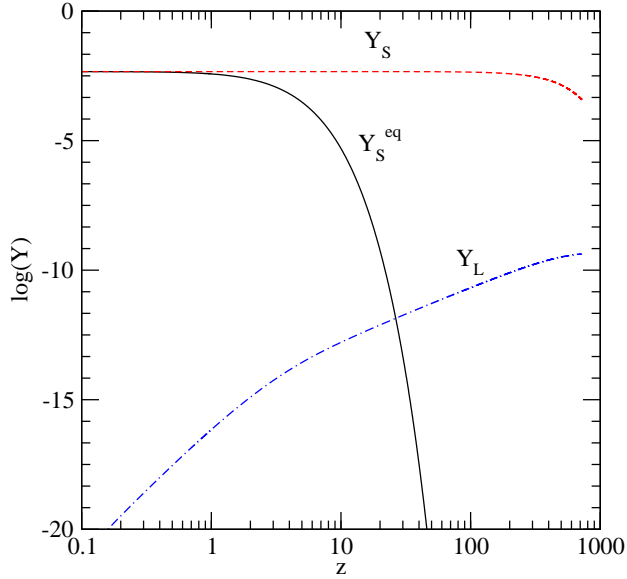}
\end{center}
\caption{The evolution of $S$ is shown against $z$ with $M_S=10^8$ GeV,
$M_1=10^9$ GeV, $M_3=10^{10}$ GeV and $d_3=10^6$ GeV and the CP asymmetry
parameter is $\epsilon_S=10^{-7}$. }
\label{d36}
\end{figure}
\begin{figure}
\begin{center}
\epsfig{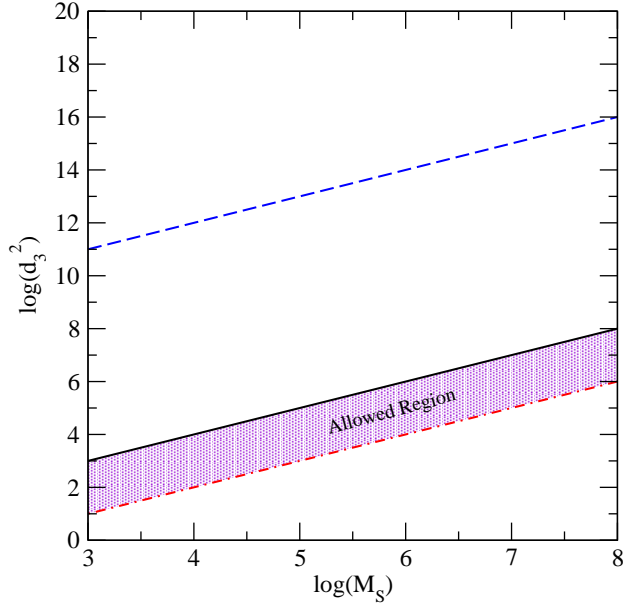}
\end{center}
\caption{The allowed values of $M_S$ are shown in the plane
of $d_3^2$ versus $M_S$.}
\label{allowed_Ms}
\end{figure}

The evolution of the number density of $S$ and the corresponding
asymmetry with respect to $z$ are shown in figs. (\ref{d38}), (\ref{d37})
and (\ref{d36}). At any epoch $z$ the value of $Y_S$ and the corresponding
asymmetry $Y_L$ can be inferred from
\begin{equation}
\frac{z}{sH(M_S)}\gamma_D^S \propto \kappa \tilde{m}_3\,.
\label{decay-suppresion}
\end{equation}
Since the decay rate of $S$ is suppressed by a factor of $\kappa$, the
asymmetry is produced at late times depending how small it is. However,
the value of $\kappa$ cannot be made indefinitely small. Because a net
$L$-asymmetry has to build up before the electroweak phase transition
which is required to be converted to the $B$-asymmetry through the 
sphaleron transitions. The final $L$-asymmetry is numerically obtained 
for three values of $d_3$
in figs. (\ref{d38}), (\ref{d37}) and (\ref{d36}). It is found
that the final $L$-asymmetry is almost same apart from a numerical
factor. This is because for the delayed decay of $S$ the wash out
effects are comparatively small. While $\kappa=10^{-2}$ and
$10^{-4}$ are used in figs. (\ref{d38}) and (\ref{d37}), it is
of $10^{-6}$ in fig. (\ref{d36}). As seen from figs. (\ref{d38}),
(\ref{d37}) and (\ref{d36}), for $\kappa=10^{-2}, 10^{-4}, 10^{-6}$
the value of $Y_L$ is saturated at around $z_s\simeq 10, 10^2, 10^3$
respectively. Assuming that a final $L$-asymmetry has to be produced
before $T_{ew}\simeq 100$ GeV the minimum tolerable value of
$\kappa=10^{-12}$ is obtained. This indicates that for $M_3=10^{10}$ GeV,
$d_3^2$ and $M_S$ can be readjusted among themselves so as to get the
suppression factor $\kappa$ ranging from $10^{-2}$ to $10^{-12}$. This
is shown in fig.(\ref{allowed_Ms}). The solid line in fig.
(\ref{allowed_Ms}) is obtained for $M_S=d_3$. The region above to that
are defined by $d_3>M_S$. So these values of $d_3$ are unnatural and
are not allowed. While the region below to the solid line are defined
by $d_3<M_S$ and hence is allowed for naturalness. Thus we see that a
wide range of $M_S$ values from $10^3$ GeV to $10^8$ GeV are allowed 
that can produce the required lepton asymmetry before the electroweak 
phase transition.

\section{Conclusions}
\noindent
We accomplished the baryogenesis via leptogenesis from the decay 
of an additional singlet $S$ in a supersymmetric extended NMSSM. 
The bound coming from the out-of-equilibrium condition could be 
evaded because the couplings of the singlets cancel out from the 
asymmetry, so the couplings could be small and can satisfy the 
out-of-equilibrium condition even at low scales. In the simplest 
seesaw models the couplings of the lightest right-handed neutrino 
could not be lowered much because that will not enable the thermal 
production of these fields. However, in the present case there is 
one additional singlet field ($\chi$) which can produce these $S$ 
fields, having large couplings to them, but itself not taking part in 
leptogenesis. 

\section*{Acknowledgments}

The work of EM was supported in part by the U.~S.~Department of Energy under 
Grant No.~DE-FG03-94ER40837.

\begin{appendix}
\section{Scattering densities}
In this appendix we give the various scattering densities that 
have been used in the Boltzmann Eqs. (\ref{S-evolution}) and 
(\ref{L-evolution}).  
\bea
\gamma_{\phi,s} &=& \frac{M_S^4 M_3 m_t^2}{64 \pi^5 v^4 z}\left( 
\frac{d_3}{M_3}\right)^2 \tilde{m}_3 \int_1^\infty dx_1 \sqrt{x_1} 
K_1(z\sqrt{x_1})\left[ 1-\frac{1}{x_1}\right]^2\\
\gamma_{\phi,t} &=& \frac{M_S^4 M_3 m_t^2}{128 \pi^5 v^4 z}\left(
\frac{d_3}{M_3}\right)^2 \tilde{m}_3 \int_1^\infty dx_1 \sqrt{x_1}
K_1(z\sqrt{x_1}) \left[ 1-\frac{1}{x_1}+\frac{1}{x_1}\ln \left(
\frac{x_1-1+y}{y}\right) \right] 
\eea
where $v$ is the vev of SM Higgs and $x_1=\frac{s}{M_S^2}$, $s$ being 
the Mandelstam variable, and $y=\frac{m_\phi^2}{M_s^2}$. 
\bea
\gamma_{N1} &=& \frac{M_1^5 M_S \tilde{m}_1^2}{128\pi^5 v^4 z}\int_0^\infty 
dx_2 \sqrt{x_2} K_1\left( z\sqrt{x_2} \frac{M_1}{M_S}\right)\nonumber\\
&&\left[ 1+\frac{1}{D_1(x_2)}+\frac{x_2}{2D_1^2(x_2)}\left \{ 1+\frac{1+x_2}
{D_1(x_2)}\right\}\ln (1+x_2)\right]\\  
\gamma_{N1,t} &=& \frac{M_1^5 M_S \tilde{m}_1^2}{128\pi^5 v^4 z}\int_0^\infty
dx_2 \sqrt{x_2} K_1\left( z\sqrt{x_2} \frac{M_1}{M_S}\right)
\left[ \frac{x_2}{x_2+1}+\frac{1}{x_2}\ln (x_2+1)\right] 
\eea
where $x_2=\frac{s}{M_1^2}$ and 
\be
\frac{1}{D_1(x_2)}=\frac{x_2-1}{(x_2-1)^2+\left( \frac{\Gamma_D^2}{M_1^2}
\right)}
\ee
\end{appendix}


\end{document}